\documentclass[journal]{IEEEtran}

\usepackage{amsmath}
\usepackage{amsfonts}
\usepackage{graphicx}
\usepackage{algorithmic}
\usepackage{algorithm}
\usepackage{color}
\usepackage{import}
\usepackage{multirow}
\usepackage{cite}

\newcommand{\eq}[1]{(\ref{#1})}
\newcommand{\fig}[1]{Fig.~\ref{#1}}
\newcommand{\figs}[2]{Figs.~\ref{#1}~and~\ref{#2}}
\newcommand{\tab}[1]{Tab.~\ref{#1}}
\newcommand{\secref}[1]{Section~\ref{#1}}
\def\abs {{\rm abs}}
\title{Distributed Power Loss Minimization in Residential Micro Grids: a Communications Perspective}
\author{Riccardo Bonetto$^\dag$, Stefano Tomasin and Michele Rossi
\thanks{$^\dag$Corresponding author: Riccardo Bonetto, e-mail: bonettor@de.unipd.it. All authors are with the Department of Information Engineering, University of Padova, Via G. Gradenigo 6/B, 35131 Padova (PD), Italy (e-mail: \{bonettor, tomasin, rossi\}@dei.unipd.it).}}
\begin{document}
\maketitle

\algsetup{
	linenosize=\small,
	linenodelimiter=.
}

\begin{abstract}
The constantly increasing number of power generation devices based on renewables is calling for a transition from the centralized control of electrical distribution grids to a distributed control scenario. In this context, distributed generators (DGs) are exploited to achieve other objectives beyond supporting loads, such as the minimization of the power losses along the distribution lines.

The aim of this work is that of designing a full-fledged system that extends existing state of the art algorithms for the distributed minimization of power losses. We take into account practical aspects such as the design of a communication and coordination protocol that is resilient to link failures and manages channel access, message delivery and DG coordination. Thus, we analyze the performance of the resulting optimization and communication scheme in terms of power loss reduction, reduction of aggregate power demand, convergence rate and resilience to communication link failures.

After that, we discuss the results of a thorough simulation campaign, obtained using topologies generated through a statistical approach that has been validated in previous research, by also assessing the performance deviation with respect to localized schemes, where the DGs are operated independently. Our results reveal that the convergence and stability performance of the selected algorithms vary greatly. However, configurations exist for which convergence is possible within five to ten communication steps and, when just $30$\% of the nodes are DGs, the aggregate power demand is roughly halved. Also, some of the considered approaches are quite robust against link failures as they still provide gains with respect to the localized solutions for failure rates as high as $50$\%.
\end{abstract}


\section{Introduction}
The traditional centralized power distribution grid is nowadays facing two important trends: the constantly increasing power demand \cite{EIA-outlook} and the worldwide diffusion of electrical power generation devices based on renewables, e.g., photovoltaic cells and wind turbines. While the former  calls for radical changes in the way the energy is generated and delivered to the final users, we note that electrical power generation is still mostly based on biofuels, fossil fuels and nuclear plants \cite{IEA-stat}. On the other hand, the installed renewable sources do not seem to be exploited as they should be, as distributed generation devices are mostly used to sell power to the energy provider by fully injecting it (through an inverter) into the grid or to fulfill the owner's power demand, without any interaction with the distribution grid, other users, or the utility provider. The limits of the fossil fuels fields, together with the inelastic nature of the demand of these goods, cause a steady price growth of the related electrical power. Moreover, nuclear-based energy production entails high costs and is subject to a diminishing public acceptance due to safety concerns. These facts are increasingly motivating the shift from traditional centralized and hierarchical power distribution grids towards smart and distributed ones~\cite{AKAGI1, BAYOD, SMART-TRENDS}. 

When communication and smart metering capabilities are added to a power grid, local generators based on renewables (also called distributed generators, DGs) can be used to enhance the grid efficiency (in terms of power distribution, reactive power compensation and frequency stability) and to relieve electricity production plants from some of the power load. In the last few years, several grid optimization techniques have been proposed \cite{PE-DISPERSED, LOAD-UNBALANCE, GRID-CONTROL-SYNC, MICRO-STOR-MAN}, each exploiting some existing communication infrastructure and relying on online smart metering procedures \cite{GRID-IMP-EST}. 

We stress that, a common approach of previous papers has been that of taking the communication infrastructure and a suitable DG coordination protocol for granted, assuming that appropriate solutions are in place and that can be exploited by the various algorithms. While this is certainly a reasonable starting point for an initial design of distributed control schemes, in the present paper, we aim at filling the gap between the electrical optimization techniques and the peculiarities of a real communication network, such as the coordination among the agents involved in the optimization. 
This is achieved through the definition and the subsequent implementation of a suitable communication protocol, for {\it powerline communication} (PLC) infrastructures~\cite{Galli, PLC-STANDARD}, that manages the channel access, the message delivery and the required DG coordination procedures. Hence, this protocol is used in conjunction with four different power distribution loss minimization techniques~\cite{LC,SurroundControl,DORPF} analyzing the performance of the resulting optimization and communication system. In particular, we explore the system behavior in terms of power loss reduction, reduction of the aggregate power demand to the distribution network, convergence rate and resilience to communication link failures. In addition, we have extended the current based surround control (CBSC) technique of~\cite{SurroundControl} by improving its DG aggregation procedure, which allows for a further reduction of distribution power losses. 

Finally, to obtain statistically valid results, as in~\cite{TOM-PAG-AI} we have used the smart grid topology generator proposed in~\cite{PAG-AI}, which is based on the {\it small world} model of~\cite{SmallWorld}. Thus, we discuss the results of a thorough simulation campaign, obtained using the aforementioned topologies, by analyzing the performance deviation with respect to localized schemes (where the DGs are operated independently). 

To summarize, the main contributions of this paper are:
\begin{itemize}
\item a communication and coordination protocol designed to operate on smart micro grids in the presence of smart nodes (DGs) with communication capabilities. This protocol ensures that the channel access is contention free, that the shortest length path is chosen in multi-hop communication scenarios and that only one node communicates at any given time (thus avoiding collisions and subsequent packet losses);
\item a new design for CBSC~\cite{SurroundControl} that permits a further reduction of the distribution power losses with respect to those obtained by the original protocol;
\item a comparative study of the performance, in terms of distribution power loss reduction and convergence rate, of four selected control techniques for micro-grids~\cite{LC,SurroundControl,DORPF}, which constitute the state of the art in terms of distributed techniques for power loss minimization;
\item a comparative study of the performance, in terms of resilience to link failures, of the four selected control techniques.
\end{itemize}

The rest of this work is structured as follows. In \secref{sec:gridmodel} we introduce the electrical grid model and
the communication network, detailing the requisites that nodes equipped with PLC transceivers should meet in order to execute the control algorithms. In \secref{sec:optmethods} we briefly describe the four optimization techniques and, for each technique, we analyze the related communication requirements. 
In \secref{sec:clustering} we discuss the way in which nodes can be aggregated in order to form optimization clusters and we propose a novel aggregation procedure that results in a performance enhancement (in terms of distribution power loss reduction) with respect to that proposed in \cite{SurroundControl}. In \secref{sec:commproc} we present a novel resilient token ring protocol, specifically designed for tree networks, so that the nodes are able to access the communication channel in a contention free manner by making sure that only one node performs the required control actions, at any given time. In \secref{sec:simsetup} we detail the simulation setup and we also describe the procedure used to generate the test networks for our simulations.
In \secref{sec:results} we present and discuss the numerical results, highlighting the most interesting features of the examined algorithms in terms of dependence on system parameters such as the link failure probability and the number of nodes that are equipped with smart functionalities. 
Finally, in \secref{sec:conclusions} we draw our final considerations.


\section{Grid Model}
\label{sec:gridmodel}

\begin{figure*}
\centering
\def\svgwidth{450pt}
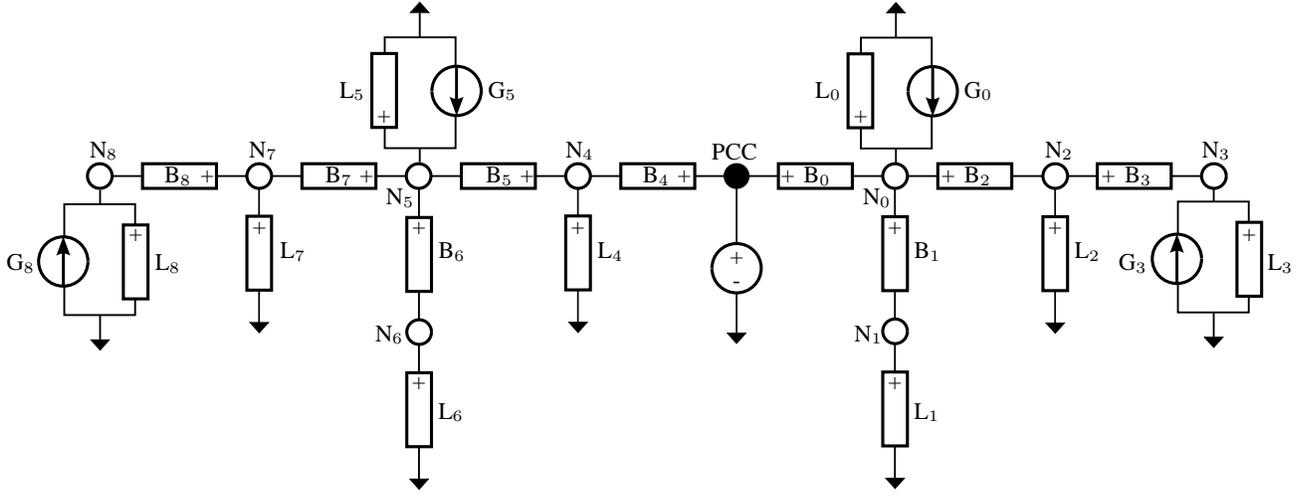
\caption{Power micro grid example.\label{figure:firstNetwork}}
\end{figure*}

We consider a power micro grid modeled as a directed tree. The root of the tree represents the point of common coupling (PCC) and the other nodes represent loads, distributed generators (DGs) and connection points. Loads are represented by complex impedances, the PCC is modeled as a voltage generator setting the voltage reference for the entire grid, while DGs are modeled as current generators.  \fig{figure:firstNetwork} shows an example of a power grid. Node $i$ is denoted by label $\text{N}_i$, load $z$ and DG $m$ are denoted respectively by $\text{L}_z$ and $\text{G}_m$  and branch impedance $j$ is denoted by $\text{B}_j$. Moreover, we assume that branches have constant impedance per meter \cite{SurroundControl, DORPF}. Note that each DG has an \textit{associated} load. This model, for example, a house (i.e., an aggregated load fed by photovoltaic panels on the rooftop). In this paper we consider the scenario where the DGs, besides feeding the respective \textit{associated} loads,  are operated in order to reduce the power distribution losses through suitable control algorithms. From \fig{figure:firstNetwork} we see that two main portions of the grid can be identified: one connecting nodes {PCC, $\text{N}_0, \dots, \text{N}_3$} (see right hand side of \fig{figure:firstNetwork}) and the other one connecting  nodes {$\text{N}_4, \dots, \text{N}_8$} (left hand side). These two portions are electrically independent and hence the DGs can be controlled separately. Generalizing this concept, if a power grid has $n\in\mathbb{N}$ branches exiting from the PCC node, then the corresponding $n$ sub-grids can be controlled in parallel. The control algorithms considered in this paper require a communication network among the controlled nodes, which is here assumed to be a powerline communication (PLC) infrastructure~\cite{Rossi-BC-2013}. Nodes equipped with PLC transceivers are referred to as smart nodes (SNs) and suitable communication protocols are assumed to allow the communication between any pair of SNs, possibly by appropriate routing of messages through intermediate SNs, as we detail shortly. We assume that SNs are also capable of measuring instantaneous electrical quantities (i.e., voltage, current and power) absorbed (by the loads) or injected (by DGs). 


\section{Optimization Methods}
\label{sec:optmethods}

In this section, four different distribution losses minimization techniques are presented and the corresponding communication requirements are discussed. One algorithm only entails localized actions and, as such, does not require any communication among nodes. This scheme will be used as a benchmark to measure the improvements that can be achieved when communication capabilities are added to the grid. 

\subsection{Local Control}
\label{ssec:LControl}

With the local control (LC) technique, \cite{LC} and \cite{OptionsLC}, each DG provides the reactive power absorbed by its \textit{associated} load. Assuming that only the reactive component injected by the inverter is controllable by the optimization process,~\cite{LC} and~\cite{OptionsLC}, LC only uses local informations available at the inverter: the active and reactive powers absorbed by the load connected to the same node. As an extension of LC, we consider the case where both the active and the reactive power generated by the DG are the powers absorbed by its \textit{associated} load and we denote this control technique as extended LC (ELC). Since LC and ELC only use local information, they do not require any communication among SNs.

\subsection{Current Based Surround Control}
\label{ssec:CBControl}

According the current based surround control (CBSC)~\cite{SurroundControl}, the grid is divided into clusters. Each cluster is composed of a pair of DGs ($\text{G}_A$ and $\text{G}_B$) such that the path connecting the DGs only contains loads. Considering a single cluster, let $I_{\text{G}_A\text{G}_B}$ be the current injected by node $\text{G}_A$ towards $\text{G}_B$ and, conversely, refer to $I_{\text{G}_B\text{G}_A}$ as the current from $\text{G}_B$ to $\text{G}_A$. The aim of CBSC is to find, for each cluster, the optimal currents $I_{\text{G}_A\text{G}_B}$ and $I_{\text{G}_B\text{G}_A}$, as we describe next. According to~\cite{SurroundControl}, fixing the initial DG, termed $\text{G}_A$, we find all the possible generators, termed $\text{G}_h$, that form a cluster with it, i.e., the portion of network between $\text{G}_A$ and $\text{G}_h$ only contains loads. $\mathcal N(\text{G}_A)$ is the set of indices of all generators $\text{G}_h$ (including the PCC) found through this procedure. Hence, the optimal current injected by $\text{G}_A$ that minimizes the distribution losses is found as:
\begin{equation}
I_{\text{G}_A}^{\text{CBSC}} = \sum_{h\in \mathcal N(\text{G}_A)}{I_{\text{G}_A\text{G}_h}^{opt}} \label{eq:CBIAopt} ,
\end{equation}
where:
\begin{equation}
	I_{\text{G}_A\text{G}_h}^{opt}
	= \displaystyle{1\over{R_{\text{G}_A,\text{G}_h}}}\sum_{i\in\mathcal{L}(\text{G}_A,\text{G}_h)}{I_{i}R_{\text{G}_h ,\text{L}_i}} \label{eq:CBIABopt}
\end{equation}
and:
\begin{itemize}
\item $R_{\text{G}_A,\text{G}_h}$ is the real part of the impedance $Z_{\text{G}_A,\text{G}_h}$ of the lines connecting DGs $\text{G}_A$ and $\text{G}_h$;
\item $R_{\text{G}_h,\text{L}_i}$ is the real part of the (total) impedance $Z_{\text{G}_h,\text{L}_i}$ of the lines connecting DG $\text{G}_h$ and load $\text{L}_i$;
\item $\mathcal{L}(\text{G}_A,\text{G}_h)$ is the set of indexes of the loads in cluster $(\text{G}_A,\text{G}_h)$;
\item $I_i,\ i\in \mathcal{L}(\text{G}_A,\text{G}_h)$ is the current absorbed by the load  $\text{L}_i$.
\end{itemize}

A variant of CBSC provides that only the reactive current injected by the DGs is controlled in order to reduce distribution losses, while the active current is regulated by other mechanisms, e.g., economic contracts or fully injected into the grid. In this case, the current injected by DG $\text{G}_A$ is $\displaystyle0+j\mathfrak{Im}(I^{\text{CBSC}}_{\text{G}_A})$.\\
In order to operate CBSC, $\text{G}_A$ first builds a list of the clusters it belongs to. This list contains the set $\mathcal N(\text{G}_A)$ and, \mbox{$\forall \, h \in \mathcal N(\text{G}_A)$}, the set $\mathcal{L}(\text{G}_A,\text{G}_h)$. Moreover, $\forall \, h\in \mathcal N(\text{G}_A)$ and $\forall \, i\in\mathcal{L}(\text{G}_A,\text{G}_h)$, $\text{G}_A$ estimates the resistances $R_{\text{G}_h\text{Li}}$ and $R_{\text{G}_A\text{G}_h}$. Once the list of clusters has been set up, Algorithm~\ref{algo:CBControlAlgo} shows the actions taken by $\text{G}_A$ in order to estimate and inject $I_{\text{G}_A}^{\text{CBSC}}$. Firstly $\text{G}_A$ creates and sends to $\text{G}_h$ a special packet denoted as DataGatheringPacket (i.e., see line~\ref{line:DGPacket}). This packet is routed to its destination by the loads whose indexes are in $\mathcal{L}(\text{G}_A,\text{G}_h)$. Once $\text{G}_h$ receives the DataGatheringPacket, it sends back an acknowledgment which, according to line~\ref{line:Ack}, is stored in the Ack variable. Each load involved in the routing process adds to the acknowledgment its index and its actual current demand. Function UpdateCluster(Ack) called in line~\ref{line:update} builds the vector CurrentDemand containing the current demands stored in the Ack variable. Elements of CurrentDemand are indexed using the indexes of $\mathcal{L}(\text{G}_A,\text{G}_h)$, as shown in line~\ref{line:SetPwDemand}. Once the optimum current has been computed, $\text{G}_A$ injects $I^{\text{CBSC}}$ as dictated by the function InjectCurrent in line~\ref{line:Inject}.

\begin{algorithm}
\caption{CBSC Pseudocode}\label{algo:CBControlAlgo}
\begin{algorithmic}[1]
\REQUIRE List of clusters
\FORALL{$h\in \mathcal N(\text{G}_A)$}
\STATE SendDataGatheringPacket(h) \label{line:DGPacket}
\STATE Ack $\leftarrow$ WaitForGatheringAck() \label{line:Ack}
\STATE CurrentDemand $\leftarrow$ UpdateCluster(Ack) \label{line:update}
\STATE $I$ $\leftarrow$ $0$ \label{line:InitCurrent}
\FORALL{$i\in\mathcal{L}(\text{G}_A,\text{G}_h)$}
\STATE $I_i$ $\leftarrow$ CurrentDemand[i] \label{line:SetPwDemand}
\STATE $\displaystyle I$ $\leftarrow$ $I+{1\over{R_{\text{G}_A\text{G}_h}}} I_i R_{\text{G}_h\text{L}_i}$ \label{line:ComputeOptC}
\ENDFOR
\STATE $I^{\text{CBSC}}$ $\leftarrow$ $I^{\text{CBSC}}+I$  \label{line:UpdateOptC}
\ENDFOR
\STATE InjectCurrent($I^{\text{CBSC}}$)  \label{line:Inject}
\end{algorithmic}
\end{algorithm}
Since the voltage reference imposed by the PCC stabilizes the grid, the current injected by the DGs does not influence the loads' total current demand. Hence CBSC requires that each DG runs Algorithm~\ref{algo:CBControlAlgo} only once in order to drive the grid's state towards the minimum distribution loss. Notably, while CBSC has very fast convergence rates, this algorithm requires that each node is a SN, which means that each node must have communication and metering capabilities (to measure or estimate the quantities required by the scheme). This may be difficult to  achieve in practice, especially when retrofitting existing grids with old equipment.

\subsection{Voltage Based Surround Control}
\label{ssec:VBControl}

The voltage based surround control (VBSC) algorithm \cite{SurroundControl} aims at reducing the communication requirements with respect to CBSC. VBSC is based on the observation that losses are minimized when all DG voltages are as close as possible to the PCC voltage. Let $\text{G}_A$ be the generator performing the control action, then the voltage that $\text{G}_A$ should reach is:
\begin{equation}
U_{\text{G}_A}^{opt} = {{\displaystyle\sum_{h\in \mathcal N(\text{G}_A)}{{R_{\text{G}_A\text{G}_h}}\over{|Z_{\text{G}_A\text{G}_h}|^2}U_{\text{G}_h}}}\over{\displaystyle\sum_{h\in \mathcal N(\text{G}_A)}{{R_{\text{G}_A\text{G}_h}}\over{|Z_{\text{G}_A\text{G}_h}|^2}}}}\label{eq:VBOptVoltage}
\end{equation}
Given \eqref{eq:VBOptVoltage}, the variation of the current injected by $\text{G}_A$ is:
\begin{equation}
\Delta I_{\text{G}_A} = {{U_{\text{G}_A}^{opt}-U_{\text{G}_A}^{0}}\over{Z^{eq}_{\text{G}_A}}}\,,\label{eq:VBOptCurrent}
\end{equation}
where $U_{\text{G}_A}^{0}$ is the actual voltage of $\text{G}_A$ and $Z^{eq}_{\text{G}_A}$ is the Thevenin impedance of the whole grid as seen by $\text{G}_A$. As for CBSC, if the active power is regulated by mechanisms other than power loss minimization, only the reactive current can be injected (see~\cite{SurroundControl}). In this case, the variation of the current injected by $\text{G}_A$ will be $0+j\mathfrak{Im}(\Delta I_{\text{G}_A})$.

\begin{algorithm}
\caption{VBSC Pseudocode}\label{algo:VBControlAlgo}
\begin{algorithmic}[1]
\REQUIRE List of neighbors $\mathcal N(\text{G}_A)$
\REQUIRE $\text{G}_A$ feeding associated load with current $I^{\text{VBSC}}$
\REQUIRE Impedance $Z_{\text{G}_A,\text{G}_h}$ $\forall \, h\in \mathcal N(\text{G}_A)$
\STATE $U^{0}_{\text{G}_A}$ $\leftarrow$ GetMyVoltage() \label{line:GetVoltage}
\FORALL{$h \in \mathcal N(\text{G}_A)$}
\STATE SendVoltageGatheringPacket($h$) \label{line:sendPkt}
\STATE Ack $\leftarrow$ WaitForGatheringAck() \label{line:rxAck}
\STATE $U_{\text{G}_h}$ $\leftarrow$ UpdateNeighborVoltage(Ack) \label{line:setNeighVoltage}
\STATE $\displaystyle{U_{num}}$ $\leftarrow$ $\displaystyle{U_{num}}+\displaystyle{\text{real}(Z_{\text{G}_A,\text{G}_h})\over{\abs(Z_{\text{G}_A,\text{G}_h})^2U_{\text{G}_h}}}$ \label{line:calcNum}
\STATE $\displaystyle{U_{den}}$ $\leftarrow$ $\displaystyle{U_{den}}+\displaystyle{\text{real}(Z_{\text{G}_A,\text{G}_h})\over{\abs(Z_{\text{G}_A,\text{G}_h})^2U_{\text{G}_h}}}$ \label{line:calcDen}
\ENDFOR
\STATE $\text{Z}^{eq}_{\text{G}_A}$ $\leftarrow$ MeasureEquivalentImpedance() \label{line:GetThImp}
\STATE $\displaystyle U^{opt}_{{\text{G}_h}}$ $\leftarrow$ $\displaystyle{U_{num}\over{U_{den}}}$ \label{line:compOptVoltage}
\STATE $\displaystyle I^{\text{VBSC}}$ $\leftarrow$ $\displaystyle I^{\text{VBSC}}+ \displaystyle{{U^{opt}_{\text{G}_h}-U^{0}_{\text{G}_A}}\over{Z^{eq}_{\text{G}_A}}}$ \label{line:updateCur}
\STATE InjectCurrent($I^{\text{VBSC}}$) \label{line:injCur}
\end{algorithmic}
\end{algorithm}

Note that the update of the current according to \eq{eq:VBOptCurrent} changes the voltages of all the other nodes, including the value of $U_{\text{G}_h},\ \forall \, \text{h} \in \mathcal N(\text{G}_A)$. Therefore, the optimum voltage is obtained through multiple control actions that gradually drive towards zero the absolute voltage difference between the DGs and the PCC. In order to operate VBSC, $\text{G}_A$ must know the indexes of the neighboring DGs ($\mathcal N(\text{G}_A)$) and the impedance of the path connecting $\text{G}_A$ and $\text{G}_h$, $\forall \, h\in \mathcal N(\text{G}_A)$. Algorithm~\ref{algo:VBControlAlgo} shows the procedure that $\text{G}_A$ executes everytime it performs the control action. According to line~\ref{line:GetVoltage},  $\text{G}_A$ firstly  stores its voltage, then for each DG whose index is in $\mathcal N(\text{G}_A)$ it sends a VoltageGatheringPacket (i.e., see line~\ref{line:sendPkt}). Once  $\text{G}_h$ receives the VoltageGatheringPacket, it measures its current voltage and sends it back to $\text{G}_A$ as an acknowledgment. Once the acknowledgment is received and the neighbor's voltage has been updated (see line~\ref{line:rxAck} and line~\ref{line:setNeighVoltage}), $\text{G}_A$ computes the numerator and denominator of \eq{eq:VBOptVoltage} corresponding to neighbor $\text{G}_h$. Once this operation has been performed for all neighboring DGs, the equivalent Thevenin impedance seen by $\text{G}_A$ (see line~\ref{line:GetThImp}) is measured and the current step is computed (see line~\ref{line:compOptVoltage} and line~\ref{line:updateCur}). Neglecting the information about the loads, this algorithm requires that only the DGs are SNs, considerably reducing the communication requirements with respect to CBSC.
\subsection{Distributed Optimal Reactive Power Flow Control}
\label{ssec:DORPFControl}
The distributed optimal reactive power flow control (DORPF) algorithm, proposed in \cite{DORPF}, assumes that only the reactive power is controlled for distributed loss minimization. DORPF groups the DGs into (possibly overlapping) clusters and, for each cluster,  a portion of the full optimization problem is solved using an approximated representation of the grid. DORPF is based on a distributed linearization of the optimal reactive power flow problem which is not reported here for the sake of conciseness (see \cite{DORPF}).  The most effective clustering strategy appears to be that of \cite{SurroundControl}, used also for CBSC. Optimizing the reactive power injection requires the generators in a cluster to estimate the PCC's voltage, the line impedance, and the neighbor's voltage. In order to get this information only distributed generators need to be SNs and hence, the communication requirements of this algorithm are similar to those of VBSC.


\section{Clustering}
\label{sec:clustering}

As discussed in the previous sections, the ability to build clusters of distributed generators is an essential feature of all the considered distributed algorithms. In this section we describe an online procedure to build clusters in a distributed fashion, by also describing a novel approach that extends the work of~\cite{SurroundControl}, and makes the optimization more robust to certain topologies. 

In~\cite{SurroundControl} and~\cite{DORPF} pairs of generators (including the PCC) such that the path connecting them only includes loads (and no generators) are defined as clusters. According to this definition, considering \fig{figure:firstNetwork}, four clusters can be identified as shown in \tab{table:GenClusters}.

\begin{table}[ht]
\caption{Generator pairs associated with the clusters in \fig{figure:firstNetwork}.}
\centering
\begin{tabular}{| c | c c |}
\hline
Cluster Number & Generator 1 & Generator 2 \\
\hline\hline
1 & PCC & G0 \\
\hline
2 & G0 & G3 \\
\hline
3 & PCC & G5 \\
\hline
4 & G5 & G8 \\
\hline
\end{tabular}
\label{table:GenClusters}
\end{table}

For VBSC and DORPF only distributed generators are required to be SNs, hence the clustering process is reduced to a neighbor discovery process. 
The CBSC algorithm, on the contrary, requires detailed informations about the loads (which are SNs too) along each cluster's path. Let $\text{G}_A$ be the DG performing the clustering procedure shown in Algorithm~\ref{algo:BClusteringG}, then to obtain this information, once $\mathcal N(\text{G}_A)$ has been set up, $\text{G}_A$ sends a special information gathering packet (called BuildClusterPacket, see line~\ref{line:buildCPkt}) to all DGs $\text{G}_h:\ h\in \mathcal N(\text{G}_A)$. Each load $\text{L}_k:\ k\in\mathcal{L}(\text{G}_A,\text{G}_h)$ appends to this packet its current demand, the impedance of the lines connecting it to $\text{G}_A$ and its identifier $k$ and forwards it to the next node in the path between $\text{G}_A$ and $\text{G}_h$. This procedure is repeated for all nodes $k$ in the path, until the BuildClusterPacket reaches the destination, as shown in Algorithm~\ref{algo:BClusteringL} (see line~\ref{line:estDist}, line~\ref{line:upPkt} and line~\ref{line:sendUpPkt}). $\text{G}_h$ stores the received load current demands and impedances from $\text{G}_A$ in the clusters table on the position corresponding to the sender's identifier and then sends back an acknowledgment with piggybacked the loads' current demands and line impedances. Once the clusters table has been set up, changes in loads current demands can be dynamically updated by the loads. 

\begin{algorithm}
\caption{Basic Clustering Pseudocode, Generator side}\label{algo:BClusteringG}
\begin{algorithmic}[1]
\REQUIRE Nodes are synchronized to central clock
\REQUIRE List of neighboring generators N \label{line:NA}
\FORALL {Neighbor n in N}
\STATE SendBuildClusterPacket(CurrentTime); \label{line:buildCPkt}
\STATE WaitForAck();
\IF {ReceivedAck()}
\STATE D$\leftarrow$SetImpVector(Ack.Impedances);
\STATE PW$\leftarrow$Ack.PowerDemand;
\STATE UpdateClusterTable(n, D, Pw);
\ENDIF
\ENDFOR
\end{algorithmic}
\end{algorithm} 

\begin{algorithm}
\caption{Basic Clustering Pseudocode, Load side}\label{algo:BClusteringL}
\begin{algorithmic}[1]
\REQUIRE Nodes are synchronized to central clock
\IF {BuildClusterPacketRx()}
\STATE d$\leftarrow$EstimateImpFromSource();\label{line:estDist}
\STATE UpdatePacket(GetPowerDemand(), d);\label{line:upPkt}
\STATE SendPacket(Packet);\label{line:sendUpPkt}
\ENDIF
\end{algorithmic}
\end{algorithm}

\begin{figure}
\centering
\includegraphics[scale=0.7]{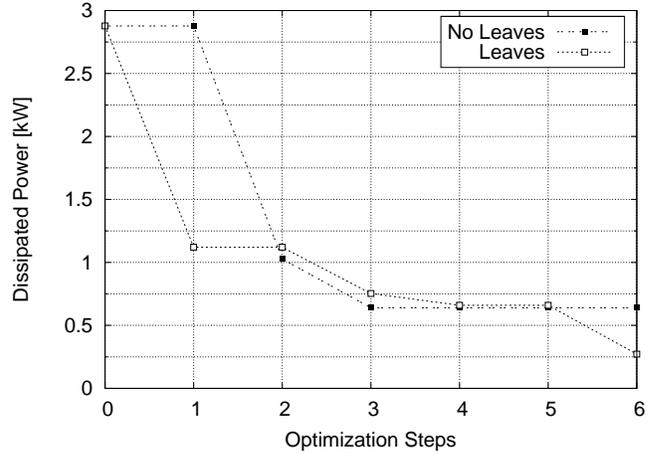}
\caption{Dissipated power {\it vs} optimization steps for CBSC for standard clustering and the proposed clustering technique.}\label{figure:compCluster}
\end{figure}

Since the CBSC requires that all the nodes are SNs, the optimization process can be improved. The clustering obtained through Algorithms \ref{algo:BClusteringG} and \ref{algo:BClusteringL} considers only couples of neighboring DGs, thus leaving out of the optimization process portions of the network ending with a leaf node with a single connected load. In \fig{figure:firstNetwork} two portions of the network are isolated: one made by edge B1 and node N1 and the other made by edge B6 and node N6. Since the leaf nodes have just one neighbor, they are able to determine their position in the network and hence to communicate to the nearest distributed generator the presence of a portion of the network that would not be optimized using the standard clustering approach. Once a DG learns of such a portion of the network, which can be achieved through a simple probing procedure, it considers it as a special cluster and fully feeds such portion of the grid. This clustering procedure (called enhanced clustering, EC) permits to enhance the performance of CBSC, with respect to the standard clustering scheme, when an appropriate number of distributed generators is accounted for. As an example, \fig{figure:compCluster} shows the performance in terms of dissipated power when EC is used. For this plot, CBSC has been executed on the topology of \fig{figure:firstNetwork} using the parameters of \tab{table:clusterParamsTable}. In \secref{ssec:impEC}, EC is further investigated for a higher number of topologies and system parameters.

\begin{table}[ht]
\caption{Parameters for the network of \fig{figure:firstNetwork}.}
\centering
\begin{tabular}{|c|ccc|}
\hline
\textbf{Active Generators} & G0 & G5 & G8\\
\hline
\textbf{Impedance [$\Omega/$m]} & \multicolumn{3}{c|}{$(0.8+j0.8)10^{-3}$}\\
\hline
\multirow{9}{*}{\textbf{Branch Lengths} [km]}
&B0&\multicolumn{2}{|c|}{0.1}\\
\cline{2-4}
&B1&\multicolumn{2}{|c|}{0.023}\\
\cline{2-4}
&B2&\multicolumn{2}{|c|}{0.045}\\
\cline{2-4}
&B3&\multicolumn{2}{|c|}{0.026}\\
\cline{2-4}
&B4&\multicolumn{2}{|c|}{0.035}\\
\cline{2-4}
&B5&\multicolumn{2}{|c|}{0.067}\\
\cline{2-4}
&B6&\multicolumn{2}{|c|}{0.032}\\
\cline{2-4}
&B7&\multicolumn{2}{|c|}{0.012}\\
\cline{2-4}
&B8&\multicolumn{2}{|c|}{0.066}\\
\hline
\end{tabular}
\label{table:clusterParamsTable}
\end{table}


\section{Communication Procedures}
\label{sec:commproc}

The update of the injected current by DGs alters the operating points of all other grid nodes. To ensure convergence of the distributed algorithms of \secref{sec:optmethods}, an iterative approach to update the injected currents has been proposed in~\cite{SurroundControl}. Nevertheless, in that paper communications aspects, such as the design of protocols, the selection of the routing paths and possible approaches to make the control resilient to link failures are not discussed. In this section, we fill this gap by proposing a failure resilient token ring protocol for tree networks. This protocol is then used in conjunction with the clustering algorithm of \secref{sec:clustering}, obtaining the final distributed algorithms that will be evaluated in \secref{sec:results}.

\subsection{Token Ring Protocol}

\begin{figure}
\centering
\includegraphics[scale=0.3]{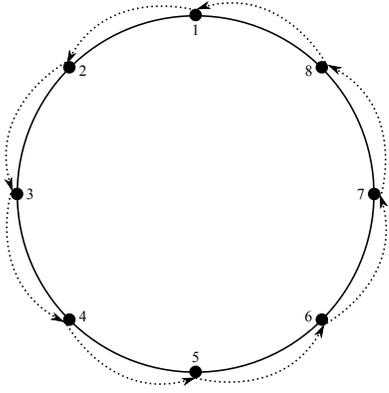}
\caption{Token ring network example. The arrows indicate the token's path. 
\label{figure:tokenRing}}
\end{figure}

As suggested in~\cite{SurroundControl}, we have implemented the selected algorithms by passing the control to a single node at a time. This is achieved through a round robin approach, whereby a {\it token} is utilized to assign the control to a certain node at a certain instant. To make sure that a single node owns the token (and hence performs the control actions as dictated by the selected algorithm) we exploit a token ring communication protocol.   

The token ring protocol (known as protocol IEEE 802.5) has originally been developed for networks whose nodes are connected in a ring fashion. Hence, a special packet called \emph{token} is circulated in the network and the node $n$ receiving the token has the right to transmit packets while all the other nodes remain silent, unless they receive a specific request from node $n$. Once the token's owner has completed its operations, it sends the token to the next neighbor, selected according to a certain schedule. \fig{figure:tokenRing} shows an example of a token ring network. Solid lines connecting the nodes numbered from $1$ to $8$ represent the actual communication links between them, while the counterclockwise pointed lines represent the token's path in the network.

\subsection{Token Ring Protocol for Tree Networks}
\label{ssec:NOS}
\begin{figure}
\centering
\includegraphics[scale=0.4]{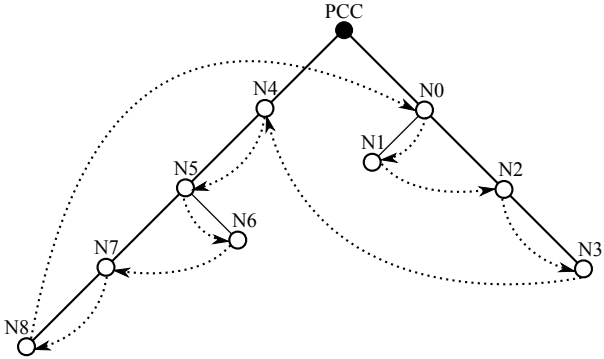}
\caption{Example of the token's path in the communication network related to the power grid of \fig{figure:firstNetwork}.\label{figure:treeTokenRing}}
\end{figure}

To adapt the standard token ring protocol to the tree topology treated in this work, a new token's owner selection procedure is devised. Let $N \geq 0$ be the number of SNs and let the current token's owner identifier be $i \in \mathbb{N}$ with $i \leq N$, then the next token's owner is attained as $j = (i+1) \, \mod \, N$. This owner selection rule ensures that when SN $i$ releases the token, all the other SNs will receive it before $i$ owns it again, thus ensuring fairness in the communication process. \fig{figure:treeTokenRing} shows an example of the token path on the power grid of \fig{figure:firstNetwork} assuming that all nodes are SNs (i.e., $N=9$). Nodes receive the token on the basis of their identifiers (on the contrary, in regular token ring networks the token exchange is based on the actual physical position of the nodes). It is also worth noting that by correctly setting the nodes' identifiers, the token's path can be forced into a depth first path search on the tree, as illustrated in \fig{figure:treeTokenRing}. Note that this minimizes the number of jumps of the token between non-adjacent nodes.

\subsection{Lost Token Recovery}

The circulating token can be lost for various reasons as, for example, external electromagnetic interference, communication link failures and device failures on the token's path. In order to recover from these events, the token reception must be acknowledged to the sender. Node $i$ releasing the token remains the owner until it successfully receives the acknowledgment sent by node $j$. If after a timeout time $T$ (also dependent on the network size and on the transmission rate) no acknowledgment is received, then the token exchange procedure is repeated until either the token reception is acknowledged or the maximum number of transmission attempts is exceeded. In the latter case, we skip the next node in the path and start a new exchange procedure with the following node $j'=(j+1) \, {\tt mod} \, N$.

\subsection{Handling Disconnected Portions of the Network}

When a portion of the network gets disconnected from a communication perspective, the nodes therein first have to discover that such an event has occurred. To this end, each SN utilizes the following approach. A timeout timer is reset at every node, every time the token is received (the timeout period must be the same for all the SNs and must be dispatched by a coordinator). If the timeout timer of a given SN counts down to zero, then the SN promotes itself as the new coordinator for the disconnected portion of the network and -from a communication standpoint- starts acting as the coordinator. It performs a new neighbor discovery, and each node involved in such process updates its routing tables according to the new information it receives. Once this process gets to completion, the obtained subtree will be optimized independently, using as a reference voltage the PCC's voltage estimated (or measured) by the new coordinator before the failure.


\section{Simulations setup}
\label{sec:simsetup}
Numerical results comparing the performance of the considered algorithms have been obtained over a large set of randomly generated grids. Instead of relying on standard test feeders as proposed in \cite{TEST-FEEDER}, we adopted the approach used in \cite{TOM-PAG-AI}. Using the generator from \cite{PAG-AI}, we generated more than a thousand power distribution grids and averaged the numerical results obtained by testing the optimization techniques taken into account on every single grid. The impact of communication and clustering protocols has also been assessed.

\subsection{Random Grid Generation}

\label{ssec:gridgeneration}
\begin{figure}
\centering
\includegraphics[scale=0.3]{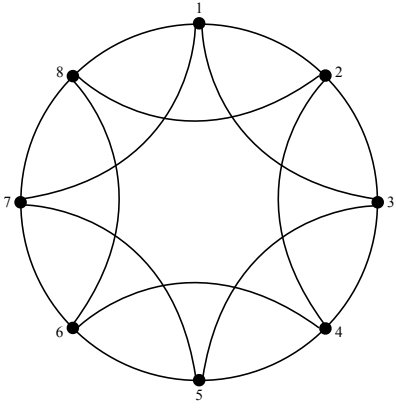}
\caption{Regular ring lattice with eight nodes.\label{figure:ringLattice}}
\end{figure}

For a meaningful performance evaluation, we have considered a large number of networks, which have been obtained using a network generator that accounts for the theoretical and experimental results of~\cite{SmallWorld}. In fact, power micro grids can be modeled as directed graphs (with the orientation of the edges determined by the direction of the active current). These graphs are included in the class of small-world networks which are connected graphs characterized by a large number of vertices with sparse connections and fill the gap between completely random graphs and regular graphs. With this approach, we can generate numerous synthetic networks by fixing their relevant parameters such as the number of nodes, the number of generators, the depth of the tree, etc., by making sure that the generated networks have statistical properties resembling those of real power grids.

\begin{figure}
\centering
\includegraphics[scale=0.3]{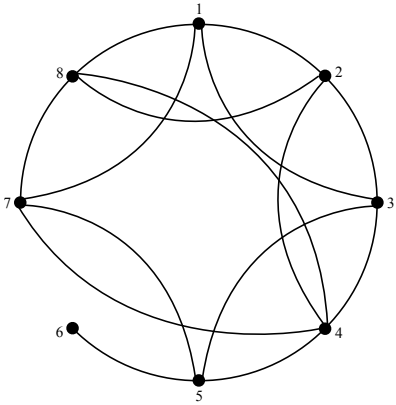}
\caption{Example of small world network generated from the regular ring lattice of \fig{figure:ringLattice}.\label{figure:smallworld}}
\end{figure}

In order to build a small-world network we start from a ring lattice. Then for each edge $e$, one of its endpoints is replaced with probability $p$ by another node, chosen uniformly at random among all other nodes. \fig{figure:ringLattice} shows an example of regular ring lattice with $8$ nodes. Instead, \fig{figure:smallworld} shows a small-world graph generated from the regular lattice of \fig{figure:ringLattice} when edges $e_a$ and $e_b$ connecting nodes $(6,7)$ and $(4,6)$ are rewired to links $(4,7)$ and $(4,8)$, respectively.

\subsection{Electrical Parameters Setup}
\label{ssec:parameters}

We assume that the generated grids operate in steady state and that they are single phased electrical networks whose distribution lines have a constant specific impedance of $(0.08+j0.08)10^{-3}$~$\Omega/\text{m}$. The phase voltage at the PCC is set at $230$~V and the voltage drops along the distribution lines are neglected in determining the loads' instantaneous power demand, as assumed in~\cite{SurroundControl}. DGs with associated loads  automatically feed them with the required current. Moreover, we do not assume limitations on the maximum current that can be injected by the DGs.

\subsection{Communication Assumptions}

\label{ssec:commassumptions}
From a communication standpoint, a first assumption is that a routing protocol connecting each pair of SNs in the grid exists. A second assumption is that no packet is lost or corrupted along a communication path unless at least one of the links in the path is broken (which is accounted for using an i.i.d. process with a certain probability).

\label{ssec:statgeneration}
In order to obtain statistically relevant results, the optimization methods treated in \secref{sec:optmethods} and the communication procedures treated in \secref{sec:commproc} and~\secref{sec:clustering} have been tested over a large number of grids and the corresponding communication network conditions. \tab{table:varyingpars} summarizes the fixed parameters considered in the simulations.

\begin{table}[ht]
\centering
\caption{Grid Parameters.}\label{table:varyingpars}
\begin{tabular}{|c|c|}
\hline
\textbf{Parameter}&\textbf{Value}\\
\hline
Number of Nodes & $30$ \\
\hline
Line Impedance & $(0.08+j0.08)10^{-3}\ \Omega/\text{m}$\\
\hline
Average Line Length & $30$~m\\
\hline
Rewiring Probability & $50\%$\\
\hline
\end{tabular}
\end{table}


\section{Numerical Results}
\label{sec:results}

In this section, we show the numerical results obtained considering the simulation setup described in \secref{sec:simsetup}.

\subsection{Dissipated Power}
\label{ssec:algoComparison}

\begin{figure}
\centering
\includegraphics[scale=0.7]{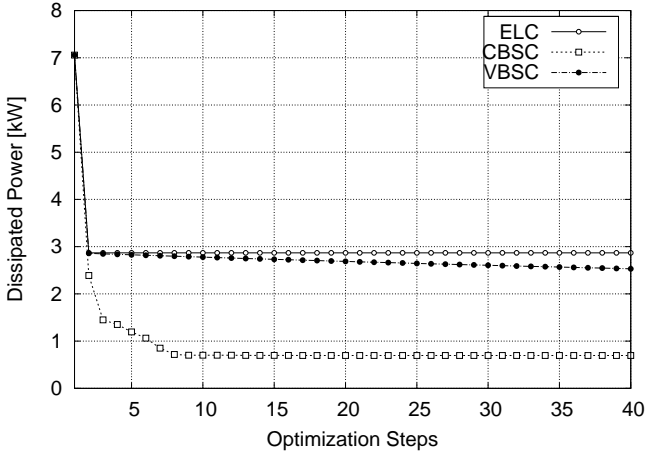}
\caption{Dissipated power {\it vs} optimization steps for ELC, CBSC and VBSC. $30$\% of nodes are DGs.\label{figure:fullPowerComp}}
\end{figure}

\begin{figure}
\centering
\includegraphics[scale=0.7]{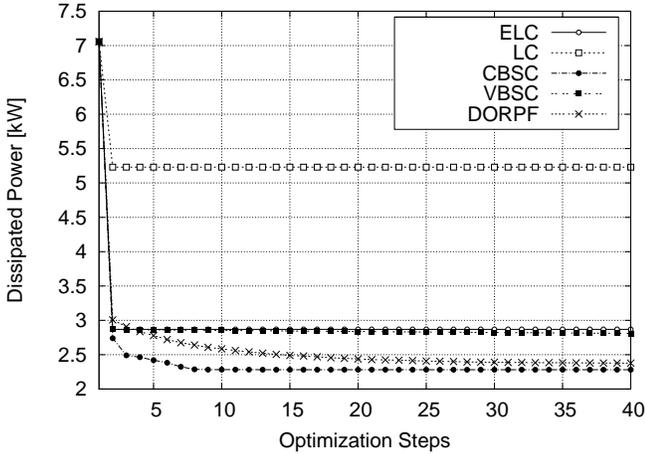}
\caption{Dissipated power {\it vs} optimization steps for LC, ELC, CBSC, VBSC and DORPF. $30$\% of nodes are DGs, reactive current injection.\label{figure:imgPowerComp}}
\end{figure}

\figs{figure:fullPowerComp}{figure:imgPowerComp} show the average dissipated power over one hundred random network realizations where both active and reactive current injection is allowed and where only reactive current injection is allowed, respectively. For each network, $30$\% of the nodes are DGs. For all algorithms, the starting point of the iterative optimization procedure provides that the DGs do not inject any current. When both active and reactive current is injected, only ELC, CBSC and VBSC are considered. A first important result that can be noticed is that both CBSC and VBSC achieve a considerably lower power distribution loss than ELC, due to their exploitation of communication capabilities. With reference to \fig{figure:fullPowerComp}, while ELC reduces the power loss by more than $7$~kW with respect to the starting point, CBSC further reduces losses by over $2$~kW in less than ten iterations. VBSC exhibits a slower convergence rate while reducing power loss by nearly $1$~kW with respect to ELC. When only reactive current is controlled (\fig{figure:imgPowerComp}), LC reduces the power losses by nearly $1.75$~kW with respect to the starting point. It is worth noting that all the distributed algorithms exploiting communication capabilities still allow the reduction of the power loss by (up to) $0.5$~kW with respect to ELC. Overall, CBSC outperforms all other approaches, relying on a complete knowledge of the clusters, since each DG has a complete knowledge of the branches connecting it to other neighboring DGs, and hence can compute the exact amount of power that is needed in each branch to minimize the loss.

\subsection{Convergence Time}
\label{ssec:algoConvergence}

\begin{figure}
\centering
\includegraphics[scale=0.7]{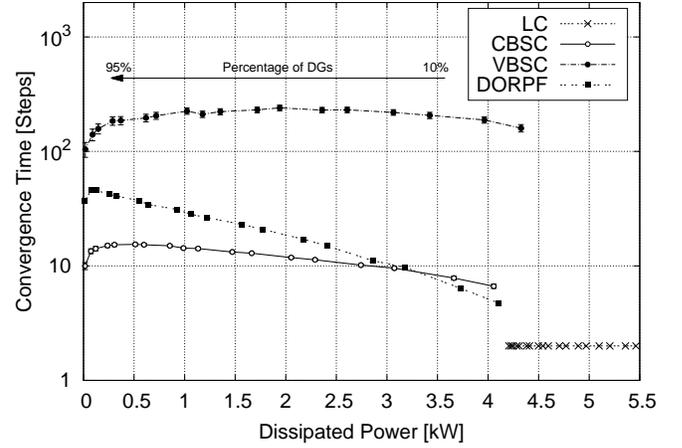}
\caption{Optimization steps {\it vs} dissipated power for ELC, CBSC, VBSC and DORPF. DGs are from $10$\% to $95$\% of nodes, reactive current injection.}\label{figure:PwT}
\end{figure}

\begin{figure}
\centering
\includegraphics[scale=0.7]{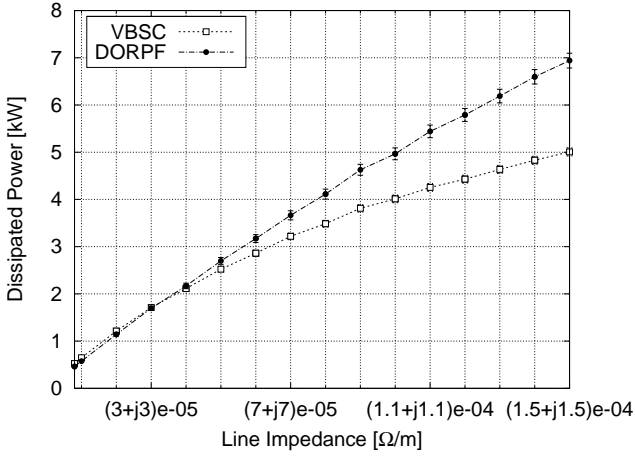}
\caption{Dissipated power {\it vs} specific line impedance for VBSC and DORPF. $15$ nodes, $50$\% rewiring probability, $30$\% of nodes are DGs, reactive current injection.\label{figure:IMPCMP}}
\end{figure}

As shown in \figs{figure:fullPowerComp}{figure:imgPowerComp}, all algorithms after a certain number of iterations converge to a minimum dissipated power. In \fig{figure:PwT}, we show the number of control steps that are required, for each algorithm, so that its performance 
falls within $5$\% of the associated minimum power loss. This number of steps is then plotted against the dissipated power by varying, as a free parameter, the percentage of DGs from $10$\% to $95$\% of the nodes. A first important result is that CBSC, VBSC and DORPF guarantee that the dissipated power is comparable to that of LC even when only $10$\% of the nodes are DGs. Moreover, the aforementioned algorithms permit to achieve a  power loss very close to zero when the DGs are about $70$\% of the nodes (or more). CBSC, once again, exhibits the fastest convergence rate and the lowest power loss at each point. It is remarkable that this algorithm always converges within a few (at most twenty for the considered networks) iterations and that this number weakly depends on the percentage of DGs. When only the distributed generators (DGs) are SNs, options reduce to DORPF and VBSC, as explained in \secref{sec:optmethods}. In this case, we note that DORPF ensures the best convergence time for the same dissipated power performance. Nevertheless, from \fig{figure:IMPCMP}, we note that when the specific line impedance grows, the  performance gap between DORPF and VBSC increases leading to a higher dissipated power, up to $25$\%, for DORPF with respect to VBSC for a specific line impedance of $(0.08+j0.08)10^{-3} \ \Omega/\text{m}$. Thus, DORPF appears to be less robust for an increasing line impedance and this fact should be carefully evaluated for practical implementations of this algorithm. In particular, when the specific line impedance is well known and ensures that the voltage drops along the power lines are less than $3$\% of the PCC's voltage, DORPF can be successfully used. On the contrary, when the specific line impedance is not known in advance or the voltage drops are not within the $3$\% range (as it may occur in rural or isolated areas), optimization should be performed through VBSC since this technique exhibits a higher robustness with respect to the grid parameters.

\subsection{Resilience to Link Failures}
\label{ssec:algoResilience}

\begin{figure}
\centering
\includegraphics[scale=0.7]{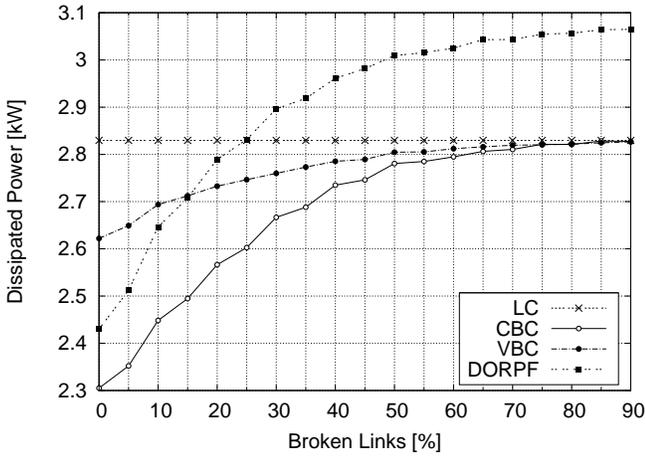}
\caption{Percentage of broken links {\it vs} dissipated power for ELC, CBSC, VBSC and DORPF. $30$\% of nodes are DGs.\label{figure:BL}}
\end{figure}

In the two previous sections we assumed a communication network with error free links. In this section, instead, the algorithms' resilience to link failures is considered. In the following results, broken links are chosen uniformly at random among all links according to a given percentage. \fig{figure:BL} shows the dissipated power as a function of the percentage of broken links; we note that despite high percentages of broken links, CBSC and VBSC achieve a lower dissipated power than ELC. On the contrary, when the percentage of broken links exceeds $25$\%, DORPF performs worse than ELC. We recall that, when only DGs are SNs only VBSC and DORPF can be used. VBSC exhibits a considerably higher degree of resilience to link failures with respect to DORPF. The higher resilience to link failures, together with the independence from the specific line impedance discussed in the previous section, make VBSC the best algorithm when little information is available about the grid or link failures are frequent events. 

\subsection{PCC Workload}

\begin{figure}
\centering
\includegraphics[scale=0.7]{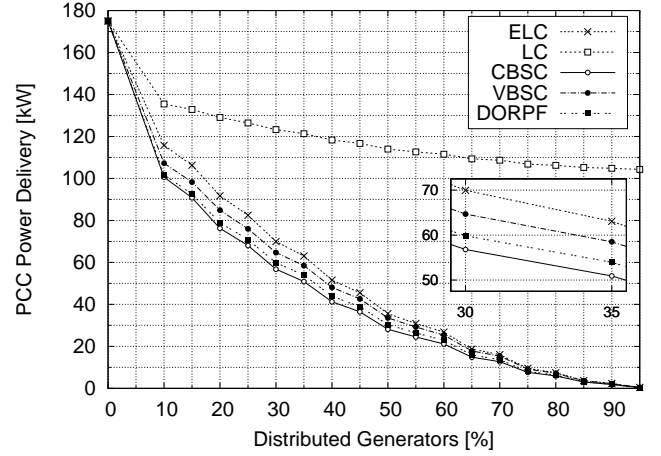}
\caption{Percentage of DGs over the total number of nodes {\it vs} PCC workload for LC, ELC, CBSC, VBSC and DORPF.\label{figure:AVGPW}}
\end{figure}

\fig{figure:AVGPW} shows the average PCC's workload as a function of the percentage of DGs in the grid. A first noticeable result is that at least $40$~kW are saved when using LC. When ELC is used, it permits to save at least $60$~kW. Distributed optimization techniques appear to be useful when the number of DGs is between $10$\% and $50$\% of the total number of nodes. In this range, the distributed optimization techniques allow to save up to $15$~kW with respect to ELC. When, instead, more than $50$\% of the nodes are DGs, the gain in terms of power loss with respect to the local control technique may not be worth the communication infrastructure needed by the distributed optimization techniques. 

\begin{figure}
\centering
\includegraphics[scale=0.7]{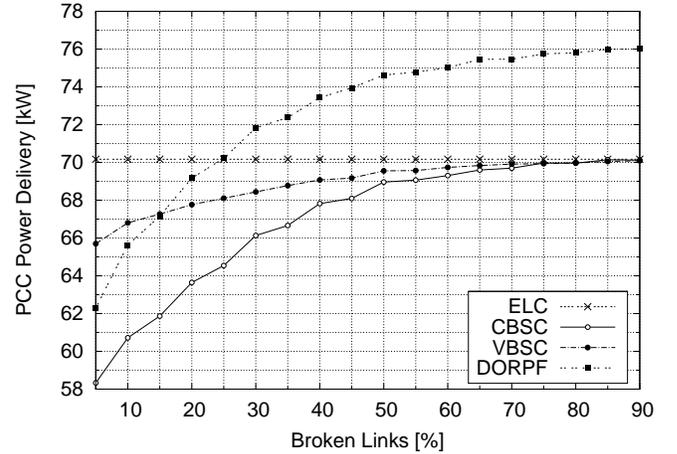}
\caption{Percentage of broken links {\it vs} PCC workload for ELC, CBSC, VBSC and DORPF. 30\% of nodes are DGs.\label{figure:AVGPWBL}}
\end{figure}

\fig{figure:AVGPWBL} shows the PCC workload as a function of the percentage of broken links in the communication network. We note that when the percentage of broken links is in the range of $10$\%-$50$\%, CBSC or VBSC outperforms the localized approach (ELC), guaranteeing a noticeable gain with respect to ELC (which, as expected, is insensitive to link failures). However, when the percentage of broken links exceeds $50$\% of the total number of links, the gain with respect to ELC is modest and may not motivate a distributed approach. Thus, when link failures can be detected, a sensible solution could be that of switching between distributed control (CBSC or VBSC) and ELC as a function of the percentage of broken links in the network.

\subsection{Impact of the EC procedure}
\label{ssec:impEC}

\begin{figure}
\centering
\includegraphics[scale=0.7]{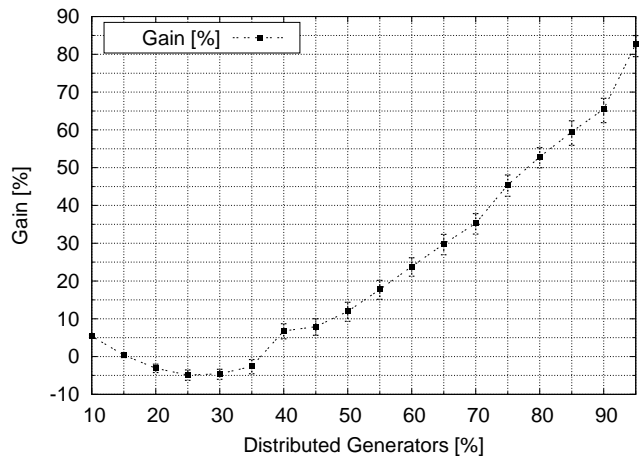}
\caption{PCC power delivery gain obtained using EC {\it vs} percentage of DGs. Reactive current injection.
\label{figure:AVGCBCPW}}
\end{figure}

In \fig{figure:AVGCBCPW} we show the relative gain (expressed as a percentage), in terms of PCC workload reduction, obtained using CBSC together with EC as opposed to using CBSC in conjunction with the standard clustering technique of \fig{figure:AVGPW}. 
In this plot we vary the percentage of nodes that are DGs from $10$\% to $95$\%. Notably, when the distributed generators are between $50$\% and $95$\% of the nodes, the gain ranges from $15$\% to $85$\%. However, in practice the cases where more than $75$\% of the nodes are DGs may be unlikely and, besides that, although the gain in this case is high in terms of percentage, the PCC workload reduction in terms of absolute value is rather small (see \fig{figure:AVGPW}). On the one hand, when DGs are between $40$\% and $75$\% of the nodes, the PCC workload reduction ranges between $3$ and $8$~kW. Note also that when the percentage of DGs is between $20$\% and $35$\%, standard clustering performs slightly better than EC. The highest gap is in this case is obtained when $30$\% of the nodes are DGs, where standard clustering permits a $5\%$ gain (i.e., about $3$~kW) with respect to EC. This is due to the fact that when the percentage of DGs is small, the special clusters optimized by EC have bigger length on average with respect to the case where the percentage of DGs is higher (i.e., above $35\%$). Hence, a higher fraction of the power fed by the DGs to the leaf nodes is wasted along the distribution lines. The PCC tries to compensate for the power dissipated on these branches injecting more reactive power and, in turn, the total power loss slightly grows.


\section{Conclusions}
\label{sec:conclusions}
In this paper, we have presented a communication-oriented design of a full-fledged system for the minimization of power losses in micro-grids, where communication and power loss optimization algorithms are jointly implemented. We have studied the performance of such system in terms of power loss reduction and convergence time when the communication network provides error free communications. We have then explored the resilience of selected optimization techniques when communication links are subject to failures. To test the impact that the communication protocol has on the optimization process, we have defined a novel contention free token ring protocol, specifically designed for tree topologies, that allows the SNs to communicate without interfering with one another. We have also devised a new aggregation procedure that permits, when the leaf nodes are SNs, a further reduction of the dissipated power that is achieved by the best algorithm. The results that we have discussed in this paper highlight the differences among the selected optimization techniques for power loss minimization. Their performance in terms of convergence time, resilience to link failures and to the specific line impedance, largely vary. However, configurations exist for which convergence is achieved within only ten communication steps. Also, the aggregate power demand of the micro-grid can be roughly halved with just $30$\% of the users being SNs.

\bibliographystyle{IEEEtran}
\bibliography{bibliography}
\nocite{*}

\enlargethispage{-45mm}

\end{document}